\documentstyle[12pt,aps,prc,epsfig,preprint]{revtex} 
\tightenlines 
\begin{document} 
\begin{titlepage} 

\title{Faddeev calculations for the $A$=5,6 $\Lambda\Lambda$ hypernuclei} 

\author{I.N. Filikhin$^{a,b}$, A. Gal$^c$ and V.M. Suslov$^a$ \\ 
$^a$Department of Mathematical and Computational Physics, 
St. Petersburg State University, 198504 Petrodvorets, 
St. Petersburg, Russia\\ 
$^b$Department of Physics, North Carolina Central University, 
Durham, NC 27707, USA\\ 
$^c$Racah Institute of Physics, The Hebrew University, 
Jerusalem 91904, Israel} 

\maketitle 

\begin{abstract} 
Faddeev calculations are reported for $^{~~5}_{\Lambda\Lambda}$H, 
$^{~~5}_{\Lambda\Lambda}$He and $^{~~6}_{\Lambda\Lambda}$He 
in terms of two $\Lambda$ hyperons plus $^3$H, $^3$He, $^4$He 
nuclear clusters respectively, using $\Lambda\Lambda$ central 
potentials considered in past non-Faddeev calculations of 
$^{~~6}_{\Lambda\Lambda}$He. The convergence with respect to the 
partial-wave expansion is studied, and comparison is made with 
some of these $\Lambda\Lambda$ hypernuclear calculations. 
The $\Lambda\Lambda - \Xi N$ mixing effect is briefly discussed. 
\newline 
\newline 
\newline 
$PACS$: 21.80.+a, 11.80.Jy, 21.45.+v, 21.30.Fe  
\newline 
{\it Keywords}: $\Lambda\Lambda$ hypernuclei; $\Lambda\Lambda$ 
interaction; Faddeev equations; few-body systems; effective interactions.  
\newline 
\newline 
Corresponding author: Avraham Gal 
\newline 
Tel: +972 2 6584930, Fax: +972 2 5611519 
\newline 
E mail: avragal@vms.huji.ac.il 
\newline 
\newline 
\centerline{(\today )} 
\end{abstract} 
\end{titlepage}

\section{Introduction} 
\label{sec:int} 

The recent report of a $^{~~6}_{\Lambda\Lambda}$He uniquely identified 
event in nuclear emulsion \cite{Tak01}, with $\Lambda \Lambda$ binding 
energy value $B_{\Lambda\Lambda} = 7.25 {\pm 0.19} ^{+0.18}_{-0.11}$ MeV, 
has triggered renewed interest in the physics of double-$\Lambda$ 
hypernuclei, particularly for light species. The previous, old report 
of $B_{\Lambda\Lambda} = 10.8 \pm 0.6$ MeV \cite{Pro66} which has been 
considered dubious by the hypernuclear community implied a fairly 
strong $\Lambda \Lambda$ interaction potential, considerably stronger 
than the $\Lambda N$ interaction potential deduced from studying 
single-$\Lambda$ hypernuclei and at odds with one-boson-exchange 
models \cite{SRi99}. In contrast, the new event is 
compatible with a fairly weak $\Lambda \Lambda$ interaction, with 
scattering length $a_{\Lambda\Lambda} \sim -0.8$ fm 
\cite{FGa02a,FGa02b} considerably smaller in magnitude than 
$a_{\Lambda N} \sim -2$ fm for the $\Lambda N$ interaction \cite{RSY99}. 
With such a weak $\Lambda \Lambda$ interaction it becomes interesting 
to explore the onset of $\Lambda \Lambda$ binding in nuclei. 
Our earlier Faddeev calculations \cite{FGa02a,FGa02b} of the $A = 5$ 
$\Lambda \Lambda$ hypernuclei $^{~~5}_{\Lambda\Lambda}$H and 
$^{~~5}_{\Lambda\Lambda}$He suggested that these species are 
comfortably particle stable for weak $\Lambda \Lambda$ interaction 
potentials, and this has been recently confirmed by the variational 
calculation of Ref. \cite{MSA02}. However, for $A = 4$ the situation 
is unclear, with conflicting calculational conclusions 
\cite{FGa02c,NAM02} for the $\Lambda \Lambda p n$ four-body 
$^{~~4}_{\Lambda \Lambda}$H hypernucleus which has been recently 
conjectured to exist in the experimental report of Ref. \cite{Ahn01}. 
A three-body ${\Lambda \Lambda N}$ bound state is ruled out on general 
grounds \cite{THe65}. 

In our earlier work \cite{FGa02a,FGa02b} the $A = 5, 6$ $\Lambda 
\Lambda$ hypernuclei $^{~~5}_{\Lambda \Lambda}$H, $^{~~5}_{\Lambda 
\Lambda}$He and $^{~~6}_{\Lambda \Lambda}$He were considered as 
three-body systems $\Lambda \Lambda C$, where the (assumed inert) 
nuclear cluster $C$ stands for $^{3}$H, $^{3}$He or $^{4}$He 
respectively. The Faddeev equations were then solved for model 
$\Lambda \Lambda$ interactions under an $s$-wave approximation 
(to be specified below) using $\Lambda C$ interaction potentials 
fitted to the observed $\Lambda C$ binding energies. 
It was argued, by comparing to earlier non-Faddeev calculations 
for $^{~~6}_{\Lambda \Lambda}$He, that the use of this 
approximation incurred an error of roughly 0.2 MeV. 
In the present work we extend our earlier Faddeev calculations, 
relaxing the $s$-wave approximation and testing the convergence 
of these calculations with respect to the partial wave expansion. 
To this end we have followed the formulation and numerical solution 
method outlined and tested by Bernab\'{e}u {\it et al.} \cite{BSS96} 
for Faddeev equations in configuration space. 
The calculations here reported do confirm our earlier estimates. 
Since $^{~~6}_{\Lambda \Lambda}$He serves in most applications as 
the primary normalizing datum for extracting phenomenologically 
the $\Lambda \Lambda$ interaction, it is desirable to improve 
as much as possible the calculational aspects of the 
$^{~~6}_{\Lambda \Lambda}$He binding energy evaluation in order 
to gain confidence in such extraction. We therefore compare 
our Faddeev calculations to other, non-Faddeev calculations 
of the $A = 5, 6$ $\Lambda \Lambda$ hypernuclei. 
Finally, we comment on the order of magnitude expected 
for dynamical effects due to $\Lambda \Lambda-\Xi N$ mixing.

\section{Methodology} 
\label{sec:meth} 

\subsection{Faddeev equations} 
\label{sec:Fad}

The bound states of the three-body systems considered in this work 
are obtained by solving the differential Faddeev equations \cite{MFa93} 

\begin{equation} 
\{H_0+V_{\alpha}(u_{\alpha})-E\} 
\Psi_{\alpha}({\bf u}_{\alpha}, {\bf v}_{\alpha}) 
= -V_{\alpha}(u_{\alpha}) 
\sum_{\beta\ne\alpha}\Psi_{\beta}({\bf u}_{\beta}, {\bf v}_{\beta}) \; \; , 
\label{1} 
\end{equation} 
where $V_\alpha$ is a short-range pair interaction in the channel $\alpha$, 
$H_0=-\Delta_{{\bf u}_{\alpha}}-\Delta_{{\bf v}_{\alpha}}$ is the internal 
kinetic energy operator, $E$ is the total energy and the wavefunction of the 
three-body system is given as a sum $\Psi =\sum^3_{\alpha=1 }\Psi_\alpha$ 
over the three Faddeev components, corresponding to the two-body 
rearrangement channels. The Faddeev components are functions of 
spin-isospin variables and of the relative Jacobi coordinate vectors, 
here given in terms of the particle coordinates 
${\bf r}_1, {\bf r}_2, {\bf r}_3$ by 

\begin{equation} 
\begin{array}{l} 
{\bf u}_{\alpha}=\left (\frac{2m_{\beta}m_{\gamma}}{m_{\beta}+m_{\gamma}} 
\right)^{1/2}({\bf r}_{\beta} - {\bf r}_{\gamma}) \;\;, \\ 
 \\ 
{\bf v}_{\alpha}=\left (\frac{2m_{\alpha}(m_{\beta}+m_{\gamma})}{M} 
\right)^{1/2}(\frac{m_{\beta}{\bf r}_{\beta}+m_{\gamma}{\bf r}_{\gamma}} 
{m_{\beta} + m_{\gamma}} - {\bf r}_{\alpha})\;\;, 
\end{array} 
\label{2} 
\end{equation} 
where ($\alpha, \beta, \gamma $) is a cyclic permutation of (1,2,3) and 
where M is the total mass. The Jacobi coordinate vectors for different 
$\alpha$'s are linearly related by an orthogonal transformation 
\begin{equation} 
  \left( 
  \begin{array}{c} 
     {\bf u}_{\alpha} \\ {\bf v}_{\alpha} 
  \end{array} 
  \right)= 
  \left( 
  \begin{array}{rl} 
      C_{\alpha\beta} & S_{\alpha\beta} \\ 
     -S_{\alpha\beta} & C_{\alpha\beta} 
  \end{array} 
  \right) 
  \left( 
  \begin{array}{c} 
     {\bf u}_{\beta} \\ {\bf v}_{\beta} 
  \end{array} 
  \right) \ ,\ \ \ C^2_{\alpha\beta} + S^2_{\alpha\beta} = 1 \;\; , 
\label{3} 
\end{equation}  
where 
\begin{equation} 
C_{\alpha\beta}=\delta_{\alpha\beta}-(1-\delta_{\alpha\beta})
\sqrt{\frac{m_{\alpha}m_{\beta}}{(M-m_{\alpha})(M-m_{\beta})}} \ \ , \ \ 
S_{\alpha\beta} = (-)^{\beta - \alpha}sgn(\beta - \alpha) 
\sqrt{1-C^{2}_{\alpha\beta}} \;\;. 
\label{4} 
\end{equation} 
The partial-wave analysis of Eq. (\ref{1}), by separating the angular 
variables (see, for instance, Ref. \cite{MGL76}), leads to a system 
of integro-differential equations, which in the polar coordinates 
$\rho^{2}={u_{\alpha}}^2 + {v_{\alpha}}^2$, 
$\tan \theta = v_{\alpha} / u_{\alpha}$, has the form 

\begin{equation} 
\begin{array}{l} 
\displaystyle  \{-\frac{\partial^{2}}{\partial \rho^{2}}-\frac {1}{\rho} 
   \frac{\partial}{\partial \rho} 
   -\frac{1}{\rho^{2}}\frac{\partial^{2}}{\partial \theta^{2}} 
   + V^{\lambda l}_{\alpha}(\rho,\theta) 
   + \frac{l(l+1)}{\rho^{2}\cos^{2}\theta} 
   + \frac{\lambda(\lambda+1)}{\rho^{2}\sin^{2}\theta} 
   - E\}\Psi^{\lambda l}_{\alpha}(\rho,\theta) = \\ 
 \\ 
\displaystyle   -\frac{1}{2}V^{\lambda l}_{\alpha}(\rho,\theta) 
\sum_{\beta\ne\alpha} ( h^{L \alpha \beta}_{\lambda l, \lambda^{'} l^{'}} 
    \Psi^{\lambda^{'} l^{'}}_{\beta})(\rho,\theta) \;\;. 
\label{5} 
\end {array} 
\end{equation} 
Here $\bf L$ is the total orbital angular momentum of the system, 
${\bf L}=\vec{\lambda}+\vec l$, where $\vec l$ is the relative 
orbital angular momentum of the pair $\alpha$ ($\alpha$=1,2,3), 
and $\vec{\lambda}$ is the orbital angular momentum of the spectator 
particle relative to the center of mass of the pair $\alpha$. 
Note that the hyper-radius $\rho$ is independent of the channel label 
$\alpha$. In a bispherical basis the integral operator has the form 
\begin{equation} 
\displaystyle (h^{L\alpha\beta}_{\lambda l, \lambda^{'} l^{'}}
   \Psi_{\beta}^{\lambda^{'} l^{'}})(\rho,\theta) = \int_{-1}^{+1}dt
   \frac{\sin\theta \cos\theta}{\sin\theta^{'} \cos\theta^{'}}
   h^{L\alpha\beta}_{\lambda l,\lambda^{'} l^{'}}(\theta,\theta^{'})
   \Psi_{\beta}^{\lambda^{'} l^{'}}(\rho,\theta^{'}) \;\; ,
\label{6} 
\end{equation} 
where 
\begin{equation} 
 \cos^{2}\theta^{'}(t,{\theta}) =  C^{2}_{\alpha\beta}\cos^{2}\theta 
  + 2 t~C_{\alpha\beta}S_{\alpha\beta}\cos\theta \sin\theta 
  + S^{2}_{\alpha\beta}\sin^{2}\theta \;\; . 
\label{7} 
\end{equation} 
An explicit representation of the operator 
$h^{L\alpha\beta}_{\lambda l,\lambda^{'} l^{'}}(\theta,\theta^{'})$ 
is given in the Appendix. 
In Eq.(\ref{5}) the potential $V^{\lambda l}_{\alpha}(\rho,\theta)$ 
has the form 
\begin{equation} 
        V^{\lambda l}_{\alpha} =<{\lambda l}|V_{\alpha}| \lambda l >  
\label{8} 
\end{equation} 
in terms of its matrix elements in the bispherical basis of eigenfunctions 
of the total angular momentum operator. The standard substitution 
$\Psi_{\alpha}=\rho^{-1/2}U_{\alpha}$ eliminates the first radial 
derivative, reducing Eq. (\ref{5}) to the form 
\begin{equation} 
\begin{array}{l} 
\displaystyle  \{-\frac{\partial^{2}}{\partial \rho^{2}} 
    -\frac{1}{\rho^{2}}\frac{\partial^{2}}{\partial \theta^{2}}
    +V^{\lambda l}_{\alpha}(\rho,\theta) 
    + \frac{l(l+1)}{\rho^{2}\cos^{2}\theta}
    +\frac{\lambda(\lambda+1)}{\rho^{2}\sin^{2}\theta}-\frac{1}{4\rho^{2}}
    - E\}U^{\lambda l}_{\alpha}(\rho,\theta) = \\
 \\
\displaystyle -\frac{1}{2}V^{\lambda l}_{\alpha}(\rho,\theta) 
      \sum_{\beta\ne\alpha} (h^{L \alpha \beta}_{\lambda l, \lambda^{'} l^{'}} 
      U^{\lambda^{'} l^{'}}_{\beta})(\rho,\theta) \;\; . 
\label{9} 
\end{array} 
\end{equation} 
To solve the eigenvalue problem in the region $\rho\in[0,\infty), \ \ \theta 
\in[0,\pi/2]$, Eq. (\ref{9}) must be supplemented by the boundary conditions 
\begin{equation}  
\begin{array}{l}
    U_{\alpha}(0,\theta) = U_{\alpha}(\infty,\theta) = 0 \;\; , \\
 \\
    U_{\alpha}(\rho,0)    = U_{\alpha}(\rho,\pi/2) = 0 \;\; . 
\label{10} 
\end{array}
\end{equation}  

\subsection{Systems with two identical particles}
\label{sec:ident} 

For a three-body system generically of the form $\Lambda \Lambda C$ 
($C$ = core), for example $\Lambda$ hyperons in 
$_{\Lambda\Lambda}^{~~6}$He ($\Lambda \Lambda \alpha$), 
the coupled set of Faddeev equations simplifies as follows: 
\begin{equation} 
\label{eq:LLC} 
\begin{array}{l} 
    (H_0+V_{\Lambda\Lambda}-E)\Psi_{C-(\Lambda\Lambda)}
   =-V_{\Lambda\Lambda}(1-P_{12})\Psi_{\Lambda-(\Lambda C)} \;\; , \\
\\
    (H_0+V_{\Lambda C}-E)\Psi_{\Lambda-(\Lambda C)}
   =-V_{\Lambda C}(\Psi_{C-(\Lambda\Lambda)}-P_{12}\Psi_{\Lambda-(\Lambda C)}) 
\;\; , 
\end{array} 
\end{equation}
where $P_{12}$ is a permutation operator for the $\Lambda$ hyperons. 
The total wavefunction is then given by 
\begin{equation} 
\label{eq:proj} 
\Psi = \Psi_{C-(\Lambda\Lambda)} + (1-P_{12})\Psi_{\Lambda-(\Lambda C)} \;\; . 
\end{equation} 
The total orbital angular momentum may be represented in two forms: 
\begin{equation} 
\label{eq:L} 
{\bf L} = {\vec \ell}_{\Lambda\Lambda}+{\vec \lambda}_{C-(\Lambda\Lambda)} 
 = {\vec \ell}_{\Lambda C}+{\vec \lambda}_{\Lambda-(\Lambda C)} \;\; . 
\end{equation} 

Other three-body systems studied in the present work are the 
isodoublet $_{\Lambda\Lambda}^{~~5}$H, $_{\Lambda\Lambda}^{~~5}$He 
charge-symmetric hypernuclei, here considered as $\Lambda\Lambda$$^3$H 
and $\Lambda\Lambda$$^3$He respectively. For the ground state 
($\frac12^+$) of these systems, after separation of spin variables 
the Faddeev equations assume the form
\begin{equation} 
\label{eq:LLCs} 
\begin{array}{l} 
   (H_0+V_{\Lambda\Lambda}-E)\Psi_{C-(\Lambda\Lambda)}
  =-V_{\Lambda\Lambda}A(1+P_{12}){\bf{\Psi}}_{\Lambda-(\Lambda C)} \;\; , \\ 
\\ 
   (H_0+V_{\Lambda C}-E){\bf{\Psi}}_{\Lambda-(\Lambda C)}=-V_{\Lambda C} 
  (A^T\Psi_{C-(\Lambda\Lambda)}+BP_{12}{\bf{\Psi}}_{\Lambda-(\Lambda C)}) 
\;\; , 
\end{array} 
\end{equation} 
where the exchange operator $P_{12}$ acts on coordinates only, 
$V_{\Lambda\Lambda}=v^s_{\Lambda\Lambda}$ is the singlet $\Lambda\Lambda$ 
potential, $V_{\Lambda C}=diag\{v^s_{\Lambda C},v^t_{\Lambda C}\}$ 
is a diagonal $2\times2$ matrix with $v^s_{\Lambda C}$ and $v^t_{\Lambda C}$ 
the singlet and triplet $\Lambda C$ interaction potentials respectively, 
and 
\begin{equation} 
\label{eq:matrix} 
A=(-\frac12,-\frac{\sqrt3}2), \qquad 
B=\left( \begin{array}{cc}
-\frac12& \frac{\sqrt3}2 \\
\frac{\sqrt3}2& \frac12 \\
\end{array} \right), \qquad 
{\bf{\Psi}}_{\Lambda-(\Lambda C)}=\left( \begin{array}{cc}
\Psi_{\Lambda-(\Lambda C)}^s \\ 
\Psi_{\Lambda-(\Lambda C)}^t \\ 
\end{array} \right).
\end{equation} 
Note that the squares of elements of $A$ correspond then to 
a $(2J+1)$ average over the $J = 0,1$ (singlet and triplet, respectively) 
$^4_\Lambda$H and $^4_\Lambda$He states. Neglecting the spin dependence 
of $V_{\Lambda C}$, i.e. using $V_{\Lambda C} = v_{\Lambda C}I$, where 
$I$ is the $2\times2$ unit matrix in spin space, it is possible to reduce 
Eqs. (\ref{eq:LLCs}) to the spinless Eqs. (\ref{eq:LLC}) where 
$\Psi_{\Lambda-(\Lambda C)}$ in Eqs. (\ref{eq:LLC}) stands for 
$A_1\Psi_{\Lambda-(\Lambda C)}^s+A_2\Psi_{\Lambda-(\Lambda C)}^t$. 
This procedure, for 
$v_{\Lambda C}=\frac{1}{4}v^s_{\Lambda C}+\frac{3}{4}v^t_{\Lambda C}$, 
will be compared below with the variational spin-averaged calculation 
of Ref. \cite{MSA02}.

\subsection{Potentials} 
\label{sec:pot} 

The $\Lambda\Lambda$ interaction potentials in the $^{1}S_0$ 
channel which are used as input to the Faddeev equations are 
of a three-range Gaussian form 
\begin{equation} 
\label{eq:HKM} 
V_{\Lambda\Lambda} = \sum_{i=1}^3 v^{(i)}\exp(-r^2/\beta_i^2)\;\;, 
\end{equation} 
following the work of Hiyama {\it et al.} \cite{HKM97} where 
a phase-equivalent $\Lambda\Lambda$ potential of this soft-core 
form was fitted to the Nijmegen model D (ND) hard-core interaction 
\cite {ND75} assuming the same hard core for the $NN$ and 
$\Lambda\Lambda$ potentials in the $^{1}S_0$ channel. 
For other interactions, notably the Nijmegen soft-core model 
NSC97 \cite{SRi99}, we have renormalized the strength of the 
medium-range attractive component ($i=2$) of this potential 
fitting as closely as possible the scattering length and 
the effective range. The appropriate range and strength parameters 
are listed in Tables 1 and 2 of Ref. \cite{FGa02b}. For the NSC97e 
interaction, Myint {\it et al.} \cite{MSA02} have used a different 
parameterization which is denoted $e$ and is listed in Table 3 of 
their paper. Finally, some older works used purely attractive 
one-range Gaussian forms, or two-range Gaussian forms with inner 
repulsion and outer attraction (this kind of potential is also 
called `Isle'). 

For the $\Lambda\alpha$ interaction potential we have followed the Isle 
potential due to Ref. \cite{KOA95} which was shown to provide good 
agreement with the measured mesonic weak decay rate of $^5_\Lambda$He. 
The resulting $\Lambda$-hyperon density distribution has been shown 
very recently \cite{NAS02} to closely resemble that due to a microscopic 
calculation of $^5_\Lambda$He using $YN$ interactions which simulate 
those of model NSC97 \cite{RSY99}. The parameters of this potential 
are listed in Table 3 of Ref. \cite{FGa02b}. Myint {\it et al.} 
\cite{MSA02} recently have used different parameters for this $s$-wave 
interaction potential, and also kept an option for using a weaker 
$p$-wave potential. These various $\Lambda\alpha$ potentials are shown 
in Fig. \ref{fig:Isle}. Similar Isle potentials were constructed for 
the $\Lambda - ^3$H and $\Lambda - ^3$He singlet and triplet 
interactions by fitting to the observed binding energies for the 
$0^+,1^+$ ground-state doublet in $^4_\Lambda$H - $^4_\Lambda$He, 
respectively \cite{FGa02b,MSA02}. Some of the older works used 
purely attractive $\Lambda C$ potentials which are no longer 
considered realistic ones. 

\section{Results and Discussion}
\label{sec:res}

In this section we report binding-energy results of solving the coupled 
Faddeev equations for $\Lambda\Lambda C$ systems with $A=5,6$, using 
a sufficiently large cutoff value $l_{\rm max}=6$ for the angular momenta 
of the partial two-body systems. Our results are compared to those of 
several non-Faddeev calculations and a brief discussion is offered. 

\subsection{$^{~~6}_{\Lambda\Lambda}$He} 
\label{sec:A=6} 

In Table \ref{tab:tabl1} we show results of our Faddeev calculations 
for $^{~~6}_{\Lambda\Lambda}$He using purely attractive Gaussian 
$\Lambda\alpha$ and $\Lambda\Lambda$ potentials taken from Ref. 
\cite{IBM85} and which act in all the allowed partial waves. 
In particular, the effective-range parameters of the $\Lambda\Lambda$ 
interaction potential are $a_{\Lambda\Lambda} = - 1.76$ fm, 
$r_{\Lambda\Lambda} = 2.11$ fm, indicating a fairly strong 
$\Lambda\Lambda$ interaction aimed at reproducing the older value for 
the binding energy $B_{\Lambda\Lambda} = 10.8 \pm 0.6$ MeV \cite{Pro66}. 
For these central interactions, the Pauli spin is conserved and 
$S=0$ generally holds for the $0^+$ ground state. Hence $L=0$ and 
$\lambda_{\alpha-(\Lambda\Lambda)}=l_{\Lambda\Lambda}$, with 
$l_{\Lambda\Lambda}$ running over even values in order to respect 
the Pauli principle. Similarly, $\lambda_{\Lambda-(\Lambda\alpha)} 
= l_{\Lambda\alpha}$. The calculated $B_{\Lambda\Lambda}$ values 
are listed in order of increasing $l_{\rm max}$, where 
$l_{\rm max}={\rm max}(l_{\Lambda\alpha},l_{\Lambda\Lambda})$, 
and are seen to increase monotonically with $l_{\rm max}$. 
Convergence is reached already for $l_{\rm max}=2$, merely 0.06 MeV 
higher than the $B_{\Lambda\Lambda}$ value corresponding to the 
$l_{\rm max}=0$ $s$-wave approximation. Our Faddeev calculations 
(marked FGS) are compared in the table to the Ikeda {\it et al.} 
\cite{IBM85} Schr\"{o}dinger-equation calculations which were restricted 
to the $\alpha-(\Lambda\Lambda)$ rearrangement channel, disregarding the 
$\Lambda-(\Lambda\alpha)$ rearrangement channel. Therefore, calculations 
of this latter type offer neither a way to sort out a range of values 
for $l_{\Lambda\alpha}$ nor a meaning for $s$-wave approximation; 
indeed, improving over what would have been perceived as an $s$-wave 
approximation ($l_{\Lambda\Lambda}=0$) amounts in Ref. \cite{IBM85} 
to 0.5 MeV, about eight times the corresponding improvement 
for our Faddeev calculation. More importantly, our Faddeev calculations 
demonstrate that the Ikeda {\it et al.} calculation misses our converged 
value of $B_{\Lambda\Lambda}$ by about 0.4 MeV (which is a sizable miss 
in this three-body trade). 

A comparison is also offered in Table \ref{tab:tabl1} with the 
variational calculation by Portilho and Coon \cite{PC91}, 
extrapolated from $l_{\rm max}=4$. The agreement with our exact Faddeev 
calculation is remarkable. This variational calculation uses a large 
basis of harmonic oscillator wavefunctions, with a variable spring 
constant, in the $\alpha-(\Lambda\Lambda)$ rearrangement channel. 
It works well because these wavefunctions within a given dimension 
may be cast into a similar basis of the same dimensionality in the 
$\Lambda-(\Lambda\alpha)$ rearrangement channel, a property which 
is exclusively specific to harmonic oscillator wavefunctions. 
Finally, we quote in the table the result of the pioneering 
Dalitz and Rajasekaran \cite{DR64} variational calculation, 
using a $^{~~6}_{\Lambda\Lambda}$He ansatz wavefunction 
\begin{equation} 
\label{eq:DR}  
\Psi = F(r_{\Lambda_1 \alpha}) F(r_{\Lambda_2 \alpha}) 
G(r_{\Lambda_1 \Lambda_2}) \;\;  
\end{equation} 
which accounts through its variational parameters for short-range 
correlations as well as for obvious long-range asymptotic requirements. 
In this outstanding calculation Dalitz and Rajasekaran used the same 
$\Lambda\Lambda$ interaction as used in the subsequent calculations 
listed in Table \ref{tab:tabl1}, but their $\Lambda\alpha$ interaction 
was slightly different, although it was constrained by fitting to the 
observed $^5_\Lambda$He binding energy. For this reason we hesitate to 
claim that their 40 years old variational calculation matches today's 
Faddeev techniques, yet even if they missed the exact result by only 
0.1 MeV (or less) it is a tribute to the essential physics 
requirements imposed on the parameterization of the functions $F$ 
and $G$ of which the wavefunction $\Psi$ in Eq. (\ref{eq:DR}) consists 
of. 

In Table \ref{tab:tabl2} we show $B_{\Lambda\Lambda}$ values calculated 
by us (marked FGS), this time using a $\Lambda\Lambda$ Isle potential 
together with a purely attractive $\Lambda\alpha$ interaction potential, 
both taken from the work of Khan and Das \cite{KD00}. These potentials, 
again, act in all the allowed partial waves. 
Our $s$-wave approximation, here too, works very well (to order of 
0.04 MeV), in contrast with the Hyperspherical Harmonics (HH) 
calculation by Khan and Das which falls short of 1 MeV in its first 
round. Nevertheless, the HH calculation is a systematical one, 
treating {\it correctly} all the three-body degrees of freedom in the 
limit of going to infinitely large values of the hyper-angular momentum 
$K$ which serves as a measure of the size of space in which the coupled 
three-body equations are solved. For $l_{\rm max}=6$ where comparison 
is possible, the agreement between the Faddeev calculation and the HH 
calculation is excellent. However, the HH calculation may suggest that 
higher values of $l_{\rm max}$ beyond $l_{\rm max}=6$ are needed in 
our Faddeev calculation in order to reach convergence, whereas this 
Faddeev calculation appears already converged at $l_{\rm max}=6$. 

\begin{figure} 
\epsfig{file=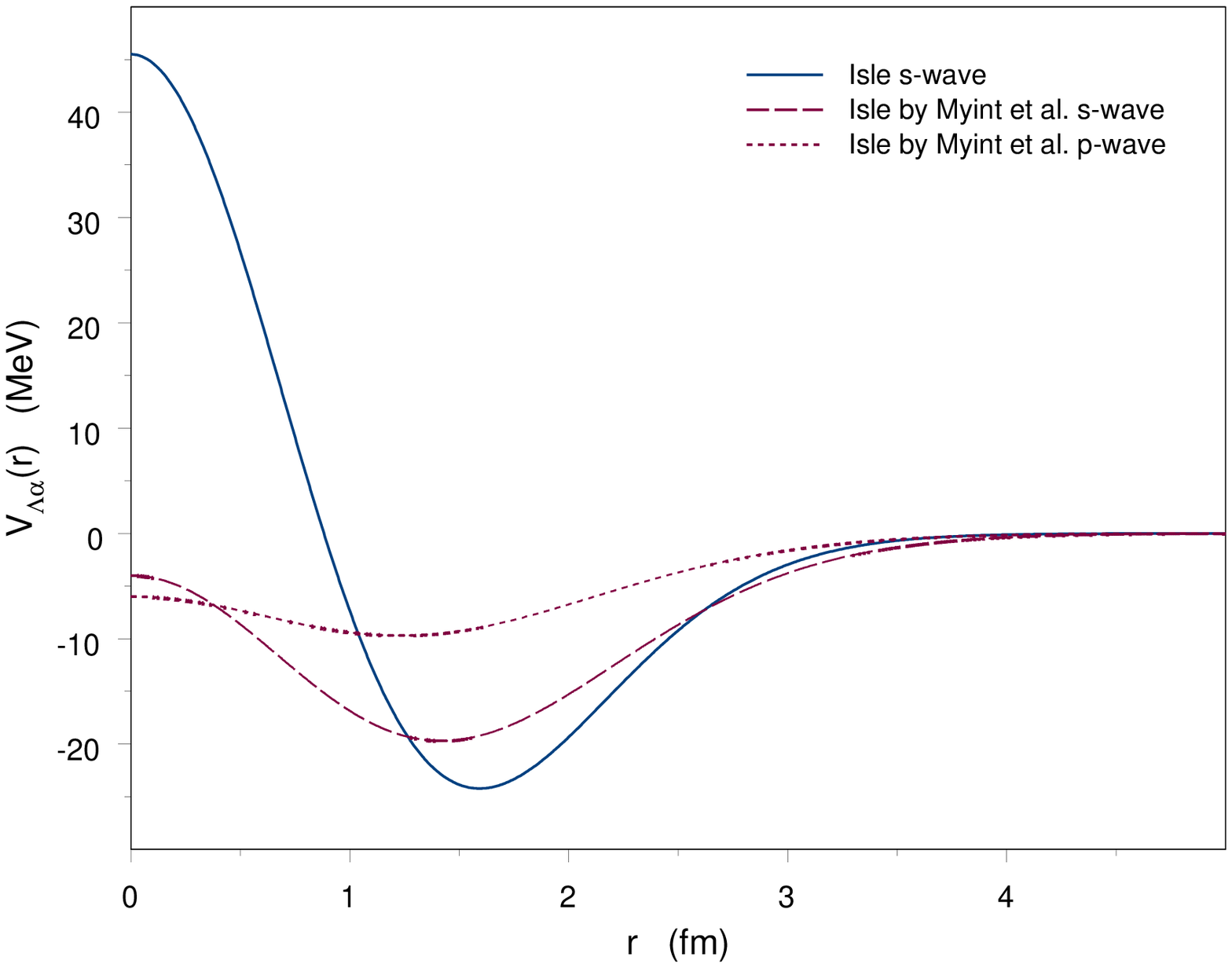, height=120mm,width=140mm} 
\vspace*{5mm} 
\caption{$\Lambda\alpha$ Isle potentials: 
solid curve from Ref. {\protect \cite{KOA95}} 
and other ones from Ref. {\protect \cite{MSA02}}.} 
\label{fig:Isle} 
\end{figure}

We now switch to a more realistic $s$-wave $\Lambda\alpha$ Isle 
interaction, with a repulsive core as shown by the solid curve 
in Fig. \ref{fig:Isle}. The $\Lambda\Lambda$ interaction which 
is of the generic form Eq. (\ref{eq:HKM}) also has a repulsive core. 
These $s$-wave interactions were recently used within an $s$-wave 
Faddeev calculation by Filikhin and Gal \cite{FGa02a,FGa02b}. Here 
we report on direct solutions of the coupled Faddeev equations for 
three specific $\Lambda \Lambda$ interaction potentials, NSC97b, 
NSC97e and ND, with parameters specified in Tables 1 and 2 of 
Ref. \cite{FGa02b}. The $\Lambda\alpha$ and $\Lambda\Lambda$ 
potentials act in all the allowed partial waves. 
The resulting $B_{\Lambda\Lambda}(^{~~6}_{\Lambda\Lambda}$He) values 
are shown in Table \ref{tab:tabl3} for $l_{\rm max} =0,1,\ldots 6$. 
It is seen that the deviation of $B_{\Lambda\Lambda}$ 
for a given value of $l_{\rm max}$ from its $s$-wave approximation 
$(l_{\rm max}=0)$ grows monotonically with $l_{\rm max}$, reaching about 
0.2 MeV for $l_{\rm max} = 6$. While it is still reasonably small, 
of the order of magnitude of the experimental error \cite{Tak01} 
derived from the observed emulsion tracks for $^{~~6}_{\Lambda\Lambda}$He, 
this deviation is several times bigger than for the calculations 
summarized in the previous tables which shared purely attractive 
$\Lambda \alpha$ potentials in common. This is due to the contribution 
of $l_{\Lambda \alpha} = 1$ getting enhanced for $\Lambda \alpha$ 
interactions of an Isle form, compared to such contribution for 
a purely attractive potential. However, since it is unrealistic 
to use the same $s$-wave $\Lambda \alpha$ interaction for all 
$\Lambda \alpha$ partial waves, one should expect the above 
deviation to be smaller once a more realistic (weaker) $p$-wave 
potential component is introduced. For example, using for 
$l_{\Lambda\alpha} = 1$ the $p$-wave Isle potential due to Myint 
{\it et al.} \cite{MSA02} which is shown by the short-dash curve 
in Fig. \ref{fig:Isle} we get within the NSC97e calculation 
$B_{\Lambda \Lambda}$ = 6.74 MeV for 
$l_{\rm max} = 1~ (l_{\Lambda\Lambda} = 0, l_{\Lambda \alpha} = 0,1$) 
compared to 6.79 MeV (cf. Table \ref{tab:tabl3}) when the standard 
$s$-wave Isle potential (solid curve in Fig. \ref{fig:Isle}) is used 
for $l_{\Lambda \alpha} = 1$ as well as for $l_{\Lambda \alpha} = 0$. 

Also shown in Table \ref{tab:tabl3} are the results of the Filikhin 
and Gal \cite{FGa02b} $s$-wave calculations which exceed by 0.11-0.15 
MeV the corresponding $l_{\rm max} = 0$ present (FGS) results. 
This discrepancy is due to the slow and non-monotonic convergence 
in the cluster-reduction method used in Refs. \cite{FGa02a,FGa02b}, 
particularly for interactions, such as here, consisting of a 
repulsive core plus an attractive tail. Finally we compare our 
Faddeev calculation with the recent variational calculation by 
Myint {\it et al.} \cite{MSA02} for the $\Lambda \Lambda$ 
interaction potential NSC97e. Their method is based on using 
Gaussian wavefunction expansion and allowing for all the 
rearrangement channels in the trial three-body wavefunction. 
The agreement between our calculation and their calculation is 
excellent both for the $s$-wave approximation ($l_{\rm max} = 0$) 
as well as for the full calculation, assuming that our Faddeev 
calculation is very close to convergence for $l_{\rm max}=6$. 

\subsection{$^{~~6}_{\Lambda\Lambda}$He and 
$^{~~5}_{\Lambda\Lambda}$H$-^{~~5}_{\Lambda \Lambda}$He}
\label{sec:A=5,6} 

In Table \ref{tab:tabl4} we show our Faddeev calculation results (FGS) 
for $^{~~6}_{\Lambda\Lambda}$He and also for the $A = 5$ charge symmetric 
$^{~~5}_{\Lambda\Lambda}$H$-^{~~5}_{\Lambda \Lambda}$He hypernuclei, 
using $s$-wave $\Lambda\alpha$ and $\Lambda\Lambda$ potentials due to 
Myint {\it et al.} \cite{MSA02}. These $s$-wave interactions are used 
in {\it all} partial waves. The results for $^{~~6}_{\Lambda \Lambda}$He 
are very similar in character to those of the previous table, indicating 
again that the $s$-wave approximation $(l_{\rm max}=0)$ holds to order 
0.2 MeV. For $^{~~5}_{\Lambda \Lambda}$H$-^{~~5}_{\Lambda \Lambda}$He 
we have a similar pattern of results, where the $s$-wave approximation 
holds to about 0.13 MeV. Here, in order to provide direct comparison 
with the variational results of Myint {\it et al.}, we have solved 
the spin-averaged form of the Faddeev equations as described here 
in Sect. \ref{sec:Fad}, using the {\it spin-averaged} $\Lambda-^{3}$H 
and $\Lambda-^{3}$He interactions listed in their paper \cite{MSA02}. 
The results of Myint {\it et al.} are given in the last row of the 
table and are in very good agreement with our $s$-wave approximation 
results. It appears that by limiting the variational calculation to 
$l_{\Lambda \alpha} = 0$, as might be understood from the discussion 
at the beginning of their section 3, the whole evaluation turned out 
to be limited to the $s$-wave approximation. We have also pursued the 
{\it full} spin-dependent calculation for the $A=5~\Lambda\Lambda$ 
hypernuclei and found out that it yields 0.08 MeV higher binding than 
for the spin-averaged results shown in Table \ref{tab:tabl4}. 

\subsection{$\Lambda\Lambda - \Xi N$ coupling effects in 
$^{~~6}_{\Lambda\Lambda}$He} 
\label{sec:mix} 

The input $\Lambda\Lambda$ interaction potentials to the Faddeev 
calculations summarized in Tables \ref{tab:tabl3},\ref{tab:tabl4} 
are {\it effective} single-channel simulations 
$V_{\Lambda\Lambda}^{\rm eff}$ of the Nijmegen meson-exchange 
models ND and NSC97 (except for $e1$ which is a slight variation 
on NSC97e). $V_{\Lambda\Lambda}^{\rm eff}$ represents the combined 
effect of the $\Lambda\Lambda$ and $\Xi N$ channels, including the 
coupling between these channels, and it is more attractive than the 
$\Lambda\Lambda$ single-channel potential $V_{\Lambda\Lambda}$ which 
does not include the effect of the $\Lambda\Lambda - \Xi N$ coupling. 
In Table \ref{tab:tabl5} we give the low-energy scattering parameters 
of, and the $^{~~6}_{\Lambda\Lambda}$He(0$^+$)  binding-energy values 
calculated by us (marked FGS) for potentials $e$ and $e1$ of Myint 
{\it et al.} \cite{MSA02}, using their parameterization of the 
corresponding potentials $V_{\Lambda\Lambda}$ and 
$V_{\Lambda\Lambda}^{\rm eff}$. It is seen that the inclusion of 
a coupling potential, motivated by the NSC97e free-space interaction, 
increases the calculated binding energy by $0.48 \pm 0.04$ MeV. 
We note that the simulation of the $\Lambda\Lambda - \Xi N$ coupling 
potential $V_{\Lambda\Lambda - \Xi N}$ in Ref. \cite{MSA02} consists 
exclusively of an attractive component, unlike the common practice 
for the $\Lambda\Lambda$ diagonal potentials, and this is likely to 
inflate the effect calculated for the $\Lambda\Lambda - \Xi N$ coupling. 
The $\Lambda\Lambda - \Xi N$ coupling potentials due to model NSC97 
involve a subtle pattern of cancellations between pseudoscalar ($K$) 
vector ($K^*$) and scalar ($\kappa$) meson-exchange contributions, 
the net result being considerably weaker than assumed by the 
parameterization in Tables 2,3 of Ref. \cite{MSA02}. Carr {\it et al.} 
\cite{CAG97}, for stronger potentials (diagonal as well as off-diagonal, 
each consisting of a short-range repulsive component 
plus a longer-range attractive component motivated by model ND), 
found that by including the $\Lambda\Lambda - \Xi N$ coupling one 
adds 0.50 MeV as shown in the table too. A consequence of their 
methodology is that for the considerably weaker NSC97 interactions, 
the total effect of including the $\Lambda\Lambda - \Xi N$ coupling 
would amount to much less, as argued recently by Afnan and Gibson 
\cite{AG03}. 

An important consideration in the discussion of the 
$\Lambda\Lambda - \Xi N$ coupling effect is the extent 
to which this coupling is Pauli suppressed. For example, 
in $^{~~6}_{\Lambda\Lambda}$He transitions creating a fifth 
nucleon in the $1s$ shell are Pauli forbidden. 
This means that $V_{\Lambda\Lambda}^{\rm eff}$ is less attractive 
in $^{~~6}_{\Lambda\Lambda}$He than in free space where it was 
derived. Myint {\it et al.} \cite{MSA02} estimated the corresponding 
suppression for potentials $e$ and $e1$ to be {\it at least} 
as large as 0.43 MeV, almost saturating the {\it total} 0.5 MeV 
$\Lambda\Lambda - \Xi N$ coupling effect calculated for these same 
potentials. This casts doubts on the validity of their method of 
estimate of the Pauli suppression effect. We note that 
the contribution due to what they perceive as the Pauli blocked 
transition is inversely proportional to the mass difference $\Delta M$ 
between the initial $^{~~6}_{\Lambda\Lambda}$He(0$^+$) ground state and 
the intermediate $^{6}_{\Xi}$He forbidden state. This mass difference 
is estimated by them to be $\Delta M \sim 32$ MeV, ignoring the 
substantial binding that a fifth nucleon in the $1s$ shell would 
have acquired in the field of the $^4$He core. Estimating this 
extra binding to be of the order of 25 MeV, the mass difference 
would reduce to merely $\Delta M \sim 7$ MeV, resulting in an 
unacceptably large Pauli suppression effect of order 2 MeV which 
exceeds substantially the total 0.5 MeV $\Lambda\Lambda - \Xi N$ 
coupling effect. The strong dependence on $\Delta M$ makes the whole 
approach questionable. Similar objections hold for the estimates given 
in Ref. \cite{MSA02} for medium effects in the $A=5$ isodoublet 
$^{~~5}_{\Lambda\Lambda}$H$-^{~~5}_{\Lambda \Lambda}$He. 
For a realistic account of the Pauli suppression effect, and other 
medium effects, the explicit introduction of the $\Xi N$ channel 
is unavoidable, as applied by Carr {\it et al.} \cite{CAG97} and by 
Yamada and Nakamoto \cite{YNa00} who used a properly defined Pauli 
suppression projection operator within a genuine coupled channel 
calculation. Carr {\it et al.} \cite{CAG97} calculated the 
suppression effect to be 0.27 MeV out of a total of 0.50 MeV, as shown 
in Table \ref{tab:tabl5}, for the considerably stronger ND interactions. 
The table also shows similar results from Ref. \cite{YNa00}. 
The Pauli-suppressed coupled-channel potential is denoted {\bf V}(ND).  

\section{Summary}
\label{sec:sum}

In this work we have studied light $\Lambda\Lambda$ hypernuclei ($A=5,6$) 
which may be described in terms of $\Lambda\Lambda C$ ($C$=cluster) 
systems and treated by solving the three-body Faddeev equations. 
Our calculations confirm the estimates made by Filikhin and Gal 
\cite{FGa02a,FGa02b} that the $s$-wave approximation 
($l_{\rm max} = 0$) works fairly well and that the contribution 
of higher partial waves is small ($< $ 0.2 MeV) if ordered according 
to increasing $l_{\rm max}$. This is not necessarily the case for 
the other, non-Faddeev methods chosen for comparison in which the 
partial-wave ordering of successive approximations is defined only 
via $l_{\Lambda \Lambda}$, irrespective of $l_{\Lambda \alpha}$. 
A direct comparison between these two classes of calculation becomes 
fully meaningful only in the limit ${\rm max}(l,\lambda)\to \infty$. 

For $^{~~6}_{\Lambda \Lambda}$He we have also studied the model 
dependence on the partial-wave composition of the $\Lambda\alpha$ 
interaction potential, in particular when weakening this interaction 
in odd states relative to the even-state strength. This model 
dependence introduces as big uncertainty into the binding-energy 
calculation as incurred by sticking to the $s$-wave approximation. 
A proper microscopic construction of the $\Lambda\alpha$ interaction potential 
leads necessarily to a nonlocal potential (e.g. Ref. \cite{HKM97}) and is 
beyond the scope and aim of the present work. 
 
For the $A=5$ $\Lambda\Lambda$ hypernuclei we have tested the accuracy of 
averaging over the spins of the $\Lambda C$ subsystems which was found to 
miss by somewhat less than 0.1 MeV the binding energy due to the full, 
spin-dependent calculation. 

Finally, we commented on the size expected for the 
$\Lambda\Lambda - \Xi N$ mixing effect in these light 
$\Lambda\Lambda$ hypernuclei, in response to unreasonable claims 
made in Ref. \cite{MSA02}. For models such as NSC97e which are 
close to describing well the $\Lambda\Lambda$ interaction as 
deduced from $B_{\Lambda\Lambda}(_{\Lambda\Lambda}^{~~6}$He) the 
$\Lambda\Lambda - \Xi N$ coupling effect should not exceed 0.2 MeV 
in $_{\Lambda\Lambda}^{~~6}$He, and a similar order of magnitude is 
expected for this and other medium effects in the $A=5$ 
$\Lambda\Lambda$ hypernuclei. 
For comparison with the better studied $S=-1$ sector we mention 
the $0^{+}-1^{+}$ binding energy difference in $_{\Lambda}^{4}{\rm He}$, 
calculated recently by Akaishi {\it et al.} \cite{AHS00} using 
a simulation of model NSC97e with and without including the powerful 
$\Lambda N - \Sigma N$ coupling which arises primarily from one-pion 
exchange. Compared to the 0.57 MeV effect of the $\Lambda N - \Sigma N$ 
coupling which these authors calculated, a considerably smaller 
effect, due to strange-meson exchange which underlies the 
$\Lambda\Lambda - \Xi N$ coupling in $\Lambda\Lambda$ hypernuclei, 
should be expected for the light $\Lambda\Lambda$ hypernuclei considered 
in the present work.  

\vspace{8mm} 
This work was partially supported by the Israel Science Foundation 
(grant 131/01). The work of I.N.F. and V.M.S. was partially supported 
by the RFFI under Grant No. 02-02-16562.

\section{Appendix}

The expression for the integral operator 
$h^{L\alpha\beta}_{\lambda l,\lambda^{'} l^{'}}$ is well known 
\cite{MGL76}. For particles of unequal masses this function 
has the form 
\begin{equation} 
\label{eq:A1} 
\begin{array}{l}
\displaystyle h^{L\alpha\beta}_{\lambda l,\lambda^{'} l^{'}}(\theta,
    \theta^{'}) = (-)^{L+l^{'}}(2\lambda^{'}+1)(2l^{'}+1)
    ((2\lambda^{'})!(2l^{'})!(2\lambda+1)(2l+1))^{1/2}  \\ 
\\
\displaystyle \times \; \sum_{\begin{array}{l}
    \lambda_{1}+\lambda_{2}=\lambda^{'} \\
     l_{1}+l_{2}=l^{'}\end{array}}
    \frac{\sin^{\lambda_{1}+l_{1}}\theta \cos^{\lambda_{2}+l_{2}}\theta}
    {\sin^{\lambda^{'}}\theta^{'} \cos^{l^{'}}\theta^{'}}
    \frac{C^{\lambda_{1}+l_{2}}_{\alpha\beta} 
    S^{\lambda_{2}+l_{1}}_{\alpha\beta}}
    {((2\lambda_{1})!(2\lambda_{2})!(2l_{1})!(2l_{2})!)^{1/2}}  \\ 
\displaystyle \times \; 
    \sum_{\lambda^{''} l^{''}}(2\lambda^{''}+1)(2l^{''}+1)   
    \left ( \begin{array}{ccc}
    \lambda_{1} & l_{1} & \lambda^{''} \\
     0          & 0     & 0
    \end{array} \right) 
    \left ( \begin{array}{ccc}
    \lambda_{2} & l_{2} & l^{''} \\
     0          & 0     & 0
    \end{array} \right) \\ 
\displaystyle \times \; 
    \sum_{k=0}(-)^{k}(2k+1)P_{k}(t)  
    \left ( \begin{array}{ccc}
     k          & \lambda^{''} & \lambda \\
     0          & 0           & 0
    \end{array} \right)   
    \left ( \begin{array}{ccc}
     k          & l^{''} & l \\
     0          & 0     & 0
    \end{array} \right) 
    \left \{ \begin{array}{ccc}
     l          & \lambda & L \\
    \lambda^{''} & l^{''}   & k
    \end{array} \right \} 
    \left \{ \begin{array}{ccc}
    \lambda_{1} & \lambda_{2} & \lambda^{'} \\
     l_{1}      &  l_{2}      &  l^{'} \\
    \lambda^{''} &  l^{''}      &  L
    \end{array} \right \} , 
\end{array} 
\end{equation} 
in terms of Legendre polynomials and $3j$, $6j$ and $9j$ symbols.
The index $k$ runs in Eq. (\ref{eq:A1}) from zero to 
$(\lambda^{'}+l^{'}+\lambda+l)/2$. 
The $C_{\alpha\beta}, S_{\alpha\beta}$ and $\cos^{2}\theta^{'}$
are defined in the main text. For zero total orbital angular momentum 
$L=0$ ($\lambda= l,\lambda^{'}= l^{'}$), all the summations in the 
expression above may be carried out to obtain a simpler expression 
of the form 
\begin{equation} 
\label{eq:A2}  
h^{\alpha\beta}_{ll^{'}}(\theta,\theta^{'}) =
(-)^{l+l^{'}}\sqrt{(2l+1)(2l^{'}+1)}P_{l}(t)P_{l^{'}}(t^{'}) \;\;\; , 
\end{equation} 
where 
\begin{equation} 
\label{eq:A3} 
\displaystyle t^{'}= \frac{-\cos(2\theta) + (C^{2}_{\alpha\beta}
-S^{2}_{\alpha\beta})\cos(2\theta^{'})}{2C_{\alpha\beta}S_{\alpha\beta}
\sin(2\theta^{'})} 
\end{equation}  
is the cosine of the angle between the vectors ${\bf u}_{\alpha}^{'}$ 
and ${\bf v}_{\alpha}^{'}$.

\newpage

\begin{table}
\caption{$B_{\Lambda\Lambda}(_{\Lambda\Lambda}^{~~6}$He) calculated 
for the purely attractive Gaussian ${\Lambda\alpha}$ and 
$\Lambda\Lambda$ interaction potentials used by Ikeda {\it et al.} 
{\protect \cite{IBM85}}.} 
\label{tab:tabl1} 
\begin{tabular}{lcccc} 
 Ref. & $l_{\rm max}$ & 
 $l_{\Lambda\alpha}$ & $l_{\Lambda\Lambda}$ 
 & $B_{\Lambda\Lambda}$ (MeV)  \\  \hline 
 FGS  &   0   & 0             & 0        & 11.15   \\ 
              &   1   & 0,1           & 0        & 11.19   \\ 
              &   2   & 0,1,2         & 0,2      & 11.21   \\ 
              &   4   & 0,1,2,3,4     & 0,2,4    & 11.21   \\ 
              &   6   & 0,1,2,3,4,5,6 & 0,2,4,6  & 11.21   \\ 
              &       &               &          &         \\   
 \cite{IBM85} & --$^a$ & -- & 0     & 10.3  \\ 
              & --$^a$ & -- & 0,2,4 & 10.8  \\ 
              &        &    &       &        \\ 
 \cite{PC91}  & --$^a$ & -- & 0,2,4 & 11.207 \\ 
              &        &    &       &        \\ 
 \cite{DR64}  &   0    & 0  &   0   & 11.2   \\ 
\end{tabular} 
$^a$Since $l_{\Lambda\alpha}$ is not assigned specific values, 
no definite value holds for 
$l_{\rm max}={\rm max}(l_{\Lambda\alpha},l_{\Lambda\Lambda})$. 
\end{table} 

\begin{table}
\caption{$B_{\Lambda\Lambda}(_{\Lambda\Lambda}^{~~6}$He) calculated 
for the purely attractive Gaussian ${\Lambda\alpha}$ and the Isle 
$\Lambda\Lambda$ interaction potentials used in model B of Khan and Das 
{\protect \cite{KD00}}.} 
\label{tab:tabl2} 
\begin{tabular}{lcccc} 
 Ref. & $l_{\rm max}$ & 
 $l_{\Lambda\alpha}$ & $l_{\Lambda\Lambda}$ & $B_{\Lambda\Lambda}$ 
 (MeV) \\  \hline 
 FGS &  0  & 0             & 0       & 10.732 \\ 
     &  6  & 0,1,2,3,4,5,6 & 0,2,4,6 & 10.770 \\ 
     &     &               &         &        \\ 
 \cite{KD00} & --$^a$ &  --  & 0            &  9.707 \\ 
             & --$^a$ &  --  & 0,2,4        & 10.687 \\ 
             & --$^a$ &  --  & 0,2,4,6      & 10.767 \\ 
             & --$^a$ &  --  & 0,2,4,6,8,10 & 10.816 \\ 
\end{tabular} 
$^a$Since $l_{\Lambda\alpha}$ is not assigned specific values, 
no definite value holds for 
$l_{\rm max}={\rm max}(l_{\Lambda\alpha},l_{\Lambda\Lambda})$.
\end{table} 

\begin{table} 
\caption{$B_{\Lambda\Lambda}(_{\Lambda\Lambda}^{~~6}$He) (in MeV) 
calculated for the Isle ${\Lambda\alpha}$ potential and for several 
simulations of Nijmegen models for the $\Lambda\Lambda$ interaction 
used by Filikhin and Gal {\protect \cite{FGa02a,FGa02b}}. 
These $s$-wave ${\Lambda\alpha}$ and $\Lambda\Lambda$ potentials 
act in all the allowed partial waves.} 
\label{tab:tabl3} 
\begin{tabular}{lcccccc} 
 Ref. & $l_{\rm max}$$^a$ & $l_{\Lambda\alpha}$ & $l_{\Lambda\Lambda}$ & 
 $B_{\Lambda\Lambda}$(NSC97b) & $B_{\Lambda\Lambda}$(NSC97e) & 
 $B_{\Lambda\Lambda}$(ND) \\ \hline 
 FGS         &0& 0             & 0       & 6.491 & 6.710 & 8.947 \\ 
             &1& 0,1           & 0       & 6.593 & 6.793 & 8.982 \\ 
             &2& 0,1,2         & 0,2     & 6.653 & 6.853 & 9.061 \\ 
             &3& 0,1,2,3       & 0,2     & 6.676 & 6.879 & 9.123 \\ 
             &4& 0,1,2,3,4     & 0,2,4   & 6.690 & 6.894 & 9.148 \\ 
             &5& 0,1,2,3,4,5   & 0,2,4   & 6.695 & 6.900 & 9.167 \\ 
             &6& 0,1,2,3,4,5,6 & 0,2,4,6 & 6.698 & 6.903 & 9.176 \\ 
             & &               &         &       &       &       \\ 
\cite{FGa02b}&0& 0             &0        & 6.60  & 6.82  & 9.10 \\ 
             & &               &         &       &       &      \\ 
\cite{MSA02} &0& 0             & 0       & --    & 6.70  & --   \\ 
             &$>$0& $>$0       &  $>$0   & --    & 6.90  & --   \\ 
\end{tabular} 
$^{a}l_{\rm max}={\rm max}(l_{\Lambda\alpha},l_{\Lambda\Lambda})$. 
\end{table} 

\begin{table} 
\caption{$B_{\Lambda\Lambda}$ (in MeV) calculated for the $A=5,6$ 
$\Lambda\Lambda$ hypernuclei using the $s$-wave Isle $\Lambda C$ 
potentials and two $\Lambda\Lambda$ potentials $e$ and $e1$ due to 
Myint {\it et al.} {\protect \cite{MSA02}}. These $s$-wave potentials 
act in all the allowed partial waves.} 
\label{tab:tabl4} 
\begin{tabular}{lccccccc} 
 Ref. & $l_{\rm max}$$^a$ & $_{\Lambda\Lambda}^{~~6}$He ($e$) & 
$_{\Lambda\Lambda}^{~~5}$He ($e$) & $_{\Lambda\Lambda}^{~~5}$H ($e$) & 
$_{\Lambda\Lambda}^{~~6}$He ($e1$) & $_{\Lambda\Lambda}^{~~5}$He ($e1$) & 
$_{\Lambda\Lambda}^{~~5}$H ($e1$)  \\  \hline 
 FGS &0 & 6.880 & 3.527 & 3.002 & 7.254 & 3.810 & 3.261  \\ 
     &1 & 6.987 & 3.608 & 3.061 & 7.341 & 3.882 & 3.311  \\ 
     &2 & 7.045 & 3.640 & 3.088 & 7.405 & 3.918 & 3.343  \\ 
     &3 & 7.078 & 3.657 & 3.103 & 7.443 & 3.939 & 3.360  \\ 
     &4 & 7.095 & 3.665 & 3.111 & 7.463 & 3.451 & 3.370  \\ 
     &5 & 7.103 & 3.669 & 3.114 & 7.473 & 3.954 & 3.374  \\ 
     &6 & 7.107 & 3.671 & 3.115 & 7.477 & 3.956 & 3.376  \\ 
     &  &       &       &       &       &       &        \\ 
 \cite{MSA02} & ? & 6.88 & 3.51 & 2.99 & 7.25 & 3.80 &3.26  \\ 
\end{tabular} 
$^{a}l_{\rm max}={\rm max}(l_{\Lambda\alpha},l_{\Lambda\Lambda})$. 
\end{table} 

\begin{table} 
\caption{Low-energy parameters (in fm) and 
$B_{\Lambda\Lambda}(_{\Lambda\Lambda}^{~~6}$He) values (in MeV) 
calculated for several interaction models depending on whether 
or not or how the $\Lambda\Lambda - \Xi N$ coupling is 
incorporated.} 
\label{tab:tabl5} 
\begin{tabular}{lcccc} 
 Ref. & $V$(model) & $a_{\Lambda\Lambda}$ & $r_{\Lambda\Lambda}$ & 
 $B_{\Lambda\Lambda}(_{\Lambda\Lambda}^{~~6}$He) \\ \hline 
 FGS &    $V_{\Lambda\Lambda}(e)$     & -0.27 & 19.07 & 6.664 \\
  & $V_{\Lambda\Lambda}^{\rm eff}(e)$ & -0.50 &  8.51 & 7.107 \\ 
     &   $V_{\Lambda\Lambda}(e1)$     & -0.43 & 10.40 & 6.964 \\ 
 & $V_{\Lambda\Lambda}^{\rm eff}(e1)$ & -0.73 &  5.59 & 7.477 \\ 
 &                                    &       &       &       \\ 
 \cite{CAG97} & $V_{\Lambda\Lambda}$(ND) & -- &   --  & 9.508 \\ 
 & $V_{\Lambda\Lambda}^{\rm eff}$(ND) & -1.91 & 3.36 & 10.007 \\ 
 & {\bf V}(ND) &--&--&  9.738 \\ 
 &                                    &       &       &       \\ 
 \cite{YNa00} & $V_{\Lambda\Lambda}$(ND) & -- &   --  & 9.4   \\ 
 & $V_{\Lambda\Lambda}^{\rm eff}$(ND) & -1.91 & 3.36 & --     \\ 
 & {\bf V}(ND) &--&--&  9.8   \\ 
\end{tabular} 
\end{table}

\begin{thebibliography}{bit99} 

\bibitem{Tak01} H. Takahashi {\it et al.}, Phys. Rev. Lett. {\bf 87}, 
212502 (2001). 

\bibitem{Pro66} D.J. Prowse, Phys. Rev. Lett. {\bf 17}, 782 (1966). 

\bibitem{SRi99} V.G.J. Stoks, and Th.A. Rijken, Phys. Rev. C {\bf 59}, 
3009 (1999). 

\bibitem{FGa02a} I.N. Filikhin, and A. Gal, Phys. Rev. C {\bf 65}, 
041001(R) (2002). 

\bibitem{FGa02b} I.N. Filikhin, and A. Gal, Nucl. Phys. A {\bf 707}, 
491 (2002). 

\bibitem{RSY99} Th.A. Rijken, V.G.J. Stoks, and Y. Yamamoto, Phys. Rev. 
C {\bf 59}, 21 (1999). 

\bibitem{MSA02} K.S. Myint, S. Shinmura, and Y. Akaishi, Eur. Phys. J. 
A {\bf 16}, 21 (2003).  

\bibitem{FGa02c} I.N. Filikhin, and A. Gal, Phys. Rev. Lett. {\bf 89}, 
172502 (2002). 

\bibitem{NAM02} H. Nemura, Y. Akaishi, and K.S. Myint, nucl-th/0211082. 

\bibitem{Ahn01} J.K. Ahn {\it et al.}, Phys. Rev. Lett. {\bf 87}, 
132504 (2001). 

\bibitem{THe65} Y.C. Tang, and R.C. Herndon, Phys. Rev. Lett. {\bf 14}, 
991 (1965). 

\bibitem{BSS96} J. Bernab\'{e}u, V.M. Suslov, T.A. Strizh, and 
S.I. Vinitsky, Hyp. Int. {\bf 101/102}, 391 (1996). 

\bibitem{MFa93} L.D. Faddeev, and S.P. Merkuriev, {\it Quantum 
Scattering Theory for Several Particle Systems} (Kluwer Academic, 
Dordrecht, 1993). 

\bibitem{MGL76} S.P. Merkuriev, C. Gignoux, and A. Laverne, 
Ann. Phys. [NY] {\bf 99}, 30 (1976). 

\bibitem{HKM97} E. Hiyama, M. Kamimura, T. Motoba, T. Yamada, 
and Y. Yamamoto, Prog. Theor. Phys. {\bf 97}, 881 (1997); Phys. Rev. C 
{\bf 66}, 024007 (2002). 

\bibitem{ND75} M.M. Nagels, T.A. Rijken, and J.J. de Swart, Phys. 
Rev. D {\bf 12}, 744 (1975); {\it ibid.} {\bf 15}, 2547 (1977). 

\bibitem{KOA95} I. Kumagai-Fuse, S. Okabe, and Y. Akaishi, 
Phys. Lett. B {\bf 345}, 386 (1995). 

\bibitem{NAS02} H. Nemura, Y. Akaishi, and Y. Suzuki, Phys. Rev. Lett. 
{\bf 89}, 142504 (2002). 

\bibitem{IBM85} K. Ikeda, H. Band\={o}, and T. Motoba, 
Prog. Theor. Phys. Suppl. {\bf 81}, 147 (1985). 

\bibitem{PC91} O. Portilho, and S.A. Coon, J. Phys. G {\bf 17}, 1375 (1991).

\bibitem{DR64} R.H. Dalitz, and G. Rajasekaran, Nucl. Phys. {\bf 50}, 
450 (1964). 

\bibitem{KD00} M.A. Khan, and T.K. Das, Fizika B (Zagreb) 
{\bf 9}, 55 (2000). 

\bibitem{CAG97} S.B. Carr, I.R. Afnan, and B.F. Gibson, Nucl. Phys. A
{\bf 625}, 143 (1997). 

\bibitem{AG03} I.R. Afnan, and B.F. Gibson, Phys. Rev. C {\bf 67}, 
017001 (2003). 

\bibitem{YNa00} T. Yamada, and C. Nakamoto, Phys. Rev. C {\bf 62}, 
034319 (2000).

\bibitem{AHS00} Y. Akaishi, T. Harada, S. Shinmura, and K.S. Myint, 
Phys. Rev. Lett. {\bf 84}, 3539 (2000). 
 
\end{thebibliography}
\end{document}